\documentclass[twoside]{article}
\usepackage[a4paper]{geometry}
\usepackage[latin1]{inputenc} 
\usepackage[T1]{fontenc} 
\usepackage{RR}
\usepackage{hyperref}
\usepackage{graphicx}
\usepackage{epsfig}
\usepackage{subfigure}
\usepackage{amsmath,amsgen,amstext,amssymb}
\usepackage{amsfonts}
\usepackage{latexsym}

\usepackage{pjnotation}

\newcommand{\beq}{\begin{equation}}
\newcommand{\deq}{\end{equation}}

\newcommand{\beqm}{\begin{equation*}}
\newcommand{\deqm}{\end{equation*}}

\newcommand{\baq}{\begin{eqnarray}}
\newcommand{\daq}{\end{eqnarray}}

\newcommand{\baqm}{\begin{eqnarray*}}
\newcommand{\daqm}{\end{eqnarray*}}

\newcommand{\eps}{\varepsilon}

\newtheorem{theorem}{Theorem}

\RRdate{March 2012}
\RRauthor{
Konstantin Avrachenkov\thanks{INRIA Sophia Antipolis, France, K.Avrachenkov@sophia.inria.fr}%
\and
Peter Jacko\thanks{BCAM -- Basque Center for Applied Mathematics, Spain, jacko@bcamath.org}%
}
\authorhead{K. Avrachenkov \& P. Jacko}
\RRtitle{Routage des Int\'er\^ets dans CCN comme le Probl\`eme de Bandit-Manchot avec des Retards}
\RRetitle{CCN Interest Forwarding Strategy as Multi-Armed Bandit Model with Delays}
\titlehead{CCN Interest Forwarding Strategy}
\RRresume{
Nous consid\'erons le routage des int\'er\^ets dans CCN (Content Centric Networking) comme le probl\`eme de bandit-manchot
avec des retards. Nous etudions le comportement transitoire des politiques :
$\eps$-greedy, tuned $\eps$-greedy et Upper Confidence Bound (UCB).
\'Etonnamment, pour tous les trois politiques on a besoin d'un tr\`es court premi\`ere phase exploratoire.
Nous d\'emontrons que l'algorithme tuned $\eps$-greedy est presque aussi bon que l'algorithme UCB,
le meilleur algorithme actuellement disponible. Nous \'etablissons la limite uniforme logarithmique pour
l'algorithme tuned $\eps$-greedy. En outre de son application immédiate au routage des int\'er\^ets dans CCN,
les nouveaux r\'esultats th\'eoriques pour le probl\`eme de bandit-manchot avec des retards repr\'esentent
des avanc\'ees importantes dans la discipline l'apprentissage automatique.
}
\RRabstract{
We consider Content Centric Network (CCN) interest forwarding problem as a Multi-Armed Bandit (MAB) problem with delays. We
investigate the transient behaviour of the $\eps$-greedy, tuned $\eps$-greedy and Upper Confidence Bound (UCB) interest
forwarding policies. Surprisingly, for all the three policies very short initial exploratory phase is needed. We demonstrate
that the tuned $\eps$-greedy algorithm is nearly as good as the UCB algorithm, the best currently available algorithm. We
prove the uniform logarithmic bound for the tuned $\eps$-greedy algorithm. In addition to its immediate application to CCN
interest forwarding, the new theoretical results for MAB problem with delays represent significant theoretical advances in
machine learning discipline.
}
\RRmotcle{Information Centric Networks, Content Centric Networks, Routage des Int\'er\^ets,
Probl\`eme du Bandit-Manchot avec des Retards}
\RRkeyword{Information Centric Networks, Content Centric Networks, Interest Forwarding,
Multi-Armed Model with Delays}
\RRprojet{Maestro}  
\RCSophia 

\begin{document}
\RRNo{7917}
\makeRR   

\section{Introduction}

There is a conceptual clash between rapidly expanding digital information dissemination and
the host-based network architecture of the current Internet. To facilitate the dissemination
of digital information, several Information-Centric Network (ICN) architectures have
been proposed: TRIAD \cite{GC01}, DONA \cite{Ketal07}, CCN/NDN \cite{Jetal09}.
Since the CCN/NDN (Content-Centric Networking / Named Data Networking) proposal appears
to be the most elaborate, we develop our contribution in the framework and within the
terminology of CCN/NDN. For the sake of brevity, we shall refer to CCN/NDN as CCN.
The main features of the ICN paradigm, and the CCN architecture in particular, are that the content
is addressed by a unique name and can have many identical cached copies. Any of such copies can be
retrieved independently of its location. The content is typically divided into several small
chunks. A chunk is also uniquely identified. A chunk of content is located and requested by forwarding
so-called interests. A user or a CCN router can forward interests to one or more neighbour
CCN routers. Clearly, if there is no bandwidth limitation the most efficient way is to
forward interests to all available neighbour routers. However, if there is a bandwidth
limitation or the interest sender has to pay for the interest or/and delivered content,
there can be better interest forwarding strategies than simple flooding.

In the present work we suggest to view the problem of optimal interest forwarding strategy
as a Multi-Armed Bandit (MAB) problem. The MAB problem is a classical problem in machine
learning discipline in which a decision maker finds an optimal balance between exploration
and exploitation efforts. Here we adopt three well known algorithms from MAB literature:
$\eps$-greedy \cite{SB98}, tuned $\eps$-greedy and UCB \cite{ACF02}.
Our study brings advances to both networking
and machine learning disciplines. We show that the MAB algorithms allow to detect the
optimal router with very small number of interests sent to sub-optimal routers. The novelty
from machine learning perspective is that we analyze the transient period of the MAB
algorithms with delays. This is a very challenging topic with hardly any results available
in the literature. In fact, we can only cite the work \cite{E88} on MAB with delay. However,
the model in \cite{E88} is different from ours and there are many restrictive assumptions.

We expect that our MAB-based mechanisms can be integrated in the Interest Control Protocol
(ICP) which regulates the pacing of interests \cite{Cetal12}.

The paper is organized as follows. In Section \ref{sec:model} we present a formal model of the problem and describe three
algorithms that we propose for CCN interest forwarding. We analyze the initial exploratory phase of these algorithms in Section
\ref{sec:exploration}, both numerically and mathematically, providing a bound and an approximation of its duration. In Section
\ref{sec:logbound} we study the exploitation phase of the tuned $\eps$-greedy algorithm and prove a logarithmic bound on the
probability of choosing a suboptimal router. Section \ref{sec:conclusion} concludes. 



\section{Model and interest forwarding strategies}
\label{sec:model}

We suppose that a CCN router or a user can forward interests to $K$ CCN neighbour routers. We consider a discrete time model. The
slot duration can be chosen equal to the minimal duration of packet generation at the MAC layer. Therefore, we assume that at each
time slot $ t \in \setT := \{ 0, 1, 2, \dots \} $ the user can send only one interest to one of $K$ CCN neighbour routers.

CCN routers reply with delays distributed according to discrete distribution functions $F_k(x)$, $k=1,...,K$, $x=1,2,...$ with
mean denoted by $\mu_k$.
Specifically, we assume that a chunk corresponding to the interest generated at the present slot and forwarded to the neighbour
router $k$ is delivered by router $k$ after a random number of slots distributed according to the distribution function $F_k(x)$.
Thus, we shall know the effect of the action taken at the time slot $t$ only at the future time slot $t+X_k(t)$,
where $X_k(t)$ is an i.i.d. random variable generated according to $F_k(x)$.

We are interested in minimizing the expected number of interests sent to sub-optimal routers, or to sub-optimal arms in
terminology of the multi-armed bandit framework \cite{SB98}. The challenging novelty of our setting with respect to the
classical multi-armed bandit problem formulation is that the cost becomes known to the decision maker with delays.
In fact, the costs are the delays.

The optimal policy in the classical setting without delay is obtained by the Gittins index rule \cite{Gittins1979}, which
breaks the combinatorial complexity of the problem by computing the Gittins index (a history-dependent function) for each
router in isolation and then simply sending the interest at every slot to the router whose current Gittins index value is
lowest. This result significantly reduces the dimensionality of the problem, but the evaluation of the Gittins index may still
be computationally tedious, especially if the index depends on the whole history, not only on the last observed state.
Moreover, the Gittins optimality result requires that the evolution of costs from routers be mutually independent, while the
algorithms described below are efficient even for dependent arms \cite{ACF02}.

Since strictly speaking optimal policy is very likely to be very complex even in the classical setting without delay,
many researchers have proposed sensible policies and shown desirable properties of such policies \cite{LR85,ACF02}.
One desirable property of the multi-armed bandit problem policy is the uniform logarithmic bound on the number of sub-optimal
arms chosen by the decision maker. We shall establish the uniform logarithmic bound for the tuned $\eps$-greedy policy in the
case of delayed information in Section~\ref{sec:logbound}.

In the present work we consider the following three algorithms: $\eps$-greedy algorithm, tuned $\eps$-greedy algorithm, and
UCB (Upper Confidence Bound) algorithm. These are the most used multi-armed bandit algorithms, and in this paper we propose
their generalizations to the setting with delayed information.

Let us formally describe each algorithm. The $\eps$-greedy algorithm is the simplest algorithm. Its main drawback is that the
expected number of sub-optimal arms grows linearly in time. A variant of $\eps$-greedy algorithm was proposed in \cite{SB98}
for Markov Decision Process models without delay.

Denote by $T_k(t)$ the total number of interests sent to router $k$ and answered up to the end of slot $t-1$, and
\begin{align*}
A_k(\tau,t) &:= 1\{\mbox{interest sent to $k$ at $\tau$} \\
&\mbox{ and answered up to the end of slot $t-1$}\}.
\end{align*}

\paragraph{Algorithm $\eps$-greedy}
\begin{enumerate}
\item {\bf Initialization:} Choose $ t_0 \in \setT $ and $\eps \in ( 0, 1 ) $. During the first $t_0$ slots keep sending interests to
    routers in round robin fashion or randomly to routers chosen according to the uniform distribution.
\item \textbf{at each time slot $t \ge t_0$ do}
\item For each router $k$, compute the average delay:
$$
\overline{X}_{k,T_k(t)}=
\frac{1}{T_k(t)} \sum_{\tau=0}^{t-1} A_k(\tau,t) X_k(\tau)
$$
\item For each router $k$, set the index:
$$
\nu_k(t)=\overline{X}_{k,T_k(t)}.
$$
\item With probability $1-\eps$ send new interest to the router with the smallest index
or with probability $\eps$ send new interest to a uniformly randomly chosen router.
\item \textbf{end for}
\end{enumerate}

The tuned $\eps$-greedy algorithm and UCB
algorithm for models without delays have been proposed and analysed in \cite{ACF02}.
Both the tuned $\eps$-greedy and UCB algorithms have logarithmic bounds on the number
of sub-optimal arms in the case of no delays \cite{ACF02}.

\paragraph{Algorithm tuned $\eps$-greedy}
\begin{enumerate}
\item {\bf Initialization:} Choose $t_0 \in \setT $ and $\eps_0 \in ( 0, t_{0} )$.
During the first $t_0$ slots keep sending interests to routers in round robin
fashion or randomly to routers chosen according to the uniform distribution.
\item \textbf{at each time slot $t \ge t_0$ do}
\item For each router $k$, compute the average delay:
$$
\overline{X}_{k,T_k(t)}=
\frac{1}{T_k(t)} \sum_{\tau=0}^{t-1} A_k(\tau,t) X_k(\tau)
$$
\item For each router $k$, set the index:
$$
\nu_k(t)=\overline{X}_{k,T_k(t)}.
$$
\item With probability $1-\eps_0/t$ send new interest to the router with the smallest index and with probability $\eps_0/t$
    send new interest to a uniformly randomly chosen router.
\item \textbf{end for}
\end{enumerate}

\paragraph{Algorithm Upper Confidence Bound (UCB)}
\begin{enumerate}
\item {\bf Initialization:} Choose $t_0 \in \setT $ and $L>0$.
During the first $t_0$ slots keep sending interests to routers in round robin
fashion or randomly to routers chosen according to the uniform distribution.
\item \textbf{at each time slot $t \ge t_0$ do}
\item For each router $k$, compute the average delay:
$$
\overline{X}_{k,T_k(t)}=
\frac{1}{T_k(t)} \sum_{\tau=0}^{t-1} A_k(\tau,t) X_k(\tau)
$$
\item For each router $k$, set the index:
$$
\nu_k(t)=\overline{X}_{k,T_k(t)}-\sqrt{\frac{L \ln(t)}{T_k(t)}}
$$
where $L$ is so-called exploration parameter.
\item Send new interest to the CCN router with the smallest index.
\item \textbf{end for}
\end{enumerate}

In our case, since we minimize the cost, we should more appropriately call this algorithm the lower confidence bound algorithm.
However, to make an explicit connection with \cite{ACF02} we shall continue to call it the UCB algorithm. In the
previous works the UCB algorithm have shown slightly better performance than the tuned $\eps$-greedy algorithm.

\begin{figure}[t]
\centering
\includegraphics[width=8cm]{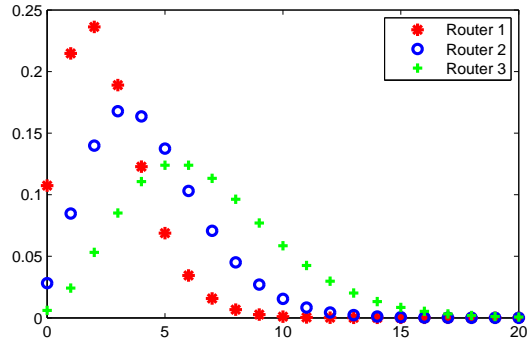}
\caption{Negative binomial distributions in example.}
\label{fig:nbdistr}
\end{figure}

To get an idea of the performance of the above algorithms in the
presence of delay, we provide a numerical example. In our numerical
examples as the distribution of delay $F_k(x)$, we have taken the
negative binomial distribution with deterministic shift. There are
several reasons for this choice. The negative binomial distribution
is quite versatile. With two parameters, we can easily choose any mean
and variance, which have simple explicit expressions. The distribution
shape can take diverse forms such as the shape of geometric distribution
and the shape close to that of the normal distribution. The negative
binomial distribution represents the distribution of a sum of
geometrically distributed random variables. Since the waiting time
distribution in many queueing systems is exponential or close to
exponential, the negative binomial distribution represents well
the response time of queueing systems in cascade. We introduce the
deterministic shift to model the propagation delay. In Table~\ref{tab:param}
we present the parameters of our numerical example and in Figure~\ref{fig:nbdistr}
we plot the negative binomial distributions with the chosen parameters.

\begin{table}[t]
\centering
\begin{tabular}{|c|c|c|c|}
  \hline
  Parameters & Router 1 & Router 2 & Router 3 \\
  \hline \hline
  propagation delay & 2 & 2 & 2 \\
  $p$ parameter & 0.8 & 0.7 & 0.6 \\
  $r$ parameter & 10 & 10 & 10 \\
  mean delay & 4.5 & 6.29 & 8.67 \\
  std & 1.77 & 2.47 & 3.33 \\
  \hline
\end{tabular}
\caption{The values of parameters in the numerical example.}
\label{tab:param}
\end{table}

\begin{figure}[t]
\centering
\includegraphics[width=8cm]{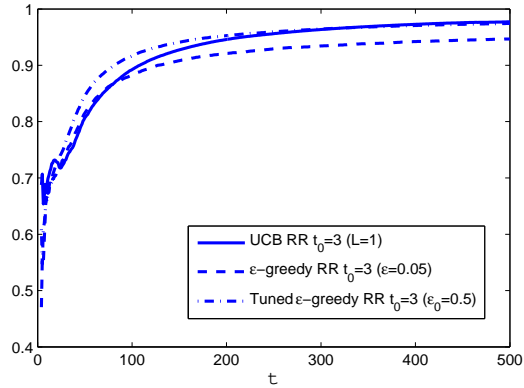}
\caption{Comparison of MAB algorithms.}
\label{fig:MABcomp1}
\end{figure}

In Figure~\ref{fig:MABcomp1} we plot the fraction of interests sent to the optimal arm as a function of time
for the three algorithms with Round Robin strategy employed in the initial phase. This numerical
example demonstrates that despite the presence of delays, the three algorithms perform well. In particular,
as in the case of no delay, the performances of the UCB and tuned $\eps$-greedy algorithms are comparable
and the $\eps$-greedy algorithm performs not too badly. In the following sections we will provide a detail
analysis of these three algorithms.

\section{Analysis of initial exploratory phase}
\label{sec:exploration}

Let us now investigate the effect of the duration of the initial, purely exploratory, phase on the algorithm performance. We
shall consider two possible initial strategies: the Round Robin (RR) strategy and the strategy when the arm chosen randomly with
uniform probability (Uni). Note that in the Round Robin strategy the initial arm and the order are chosen randomly with
uniform distribution.

\begin{figure}[t]
\centering
\includegraphics[width=8cm]{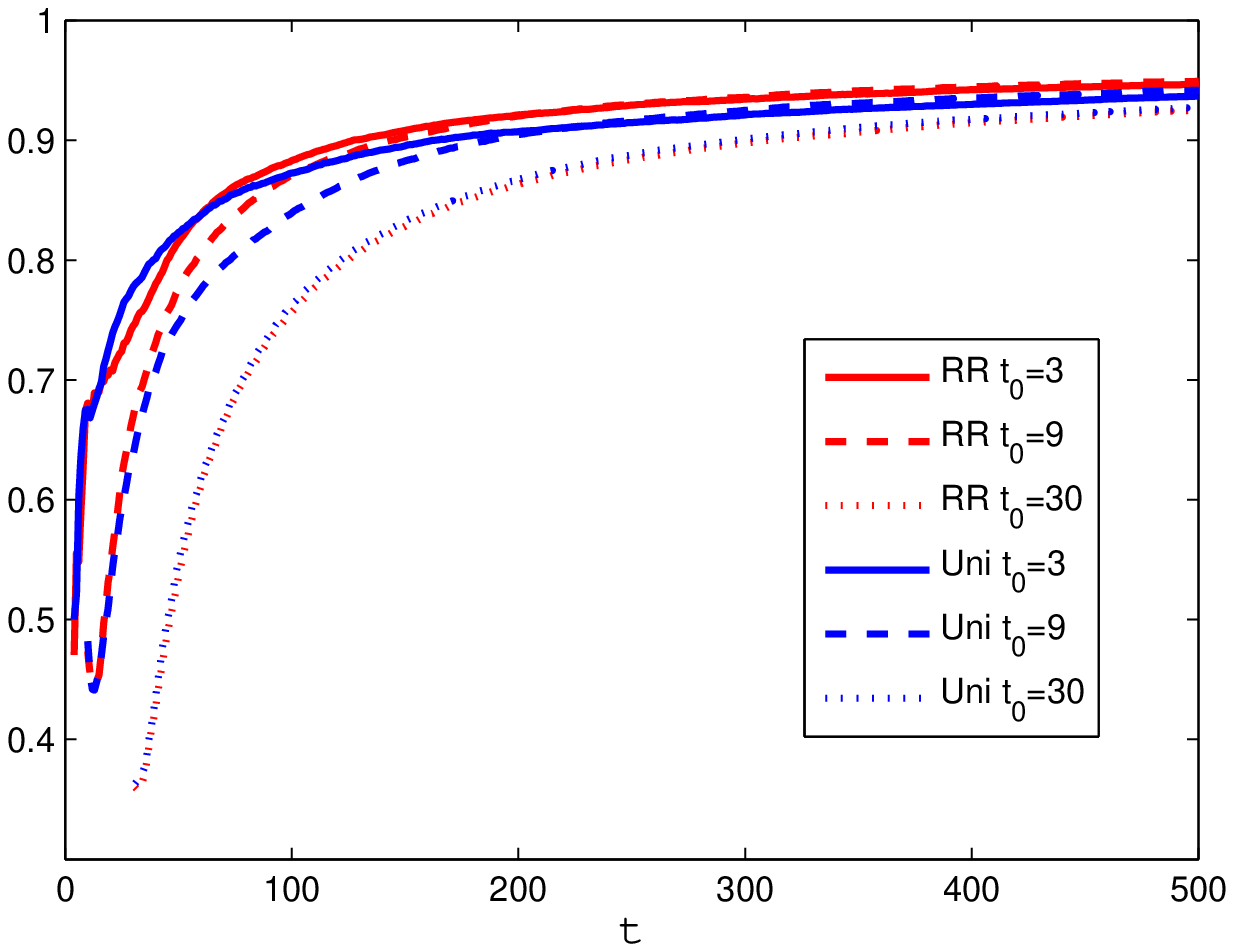}
\caption{The effect of the initial phase duration and initial strategy: $\eps$-greedy algorithm.}
\label{fig:greedyt0}
\end{figure}

\begin{figure}[t]
\centering
\includegraphics[width=8cm]{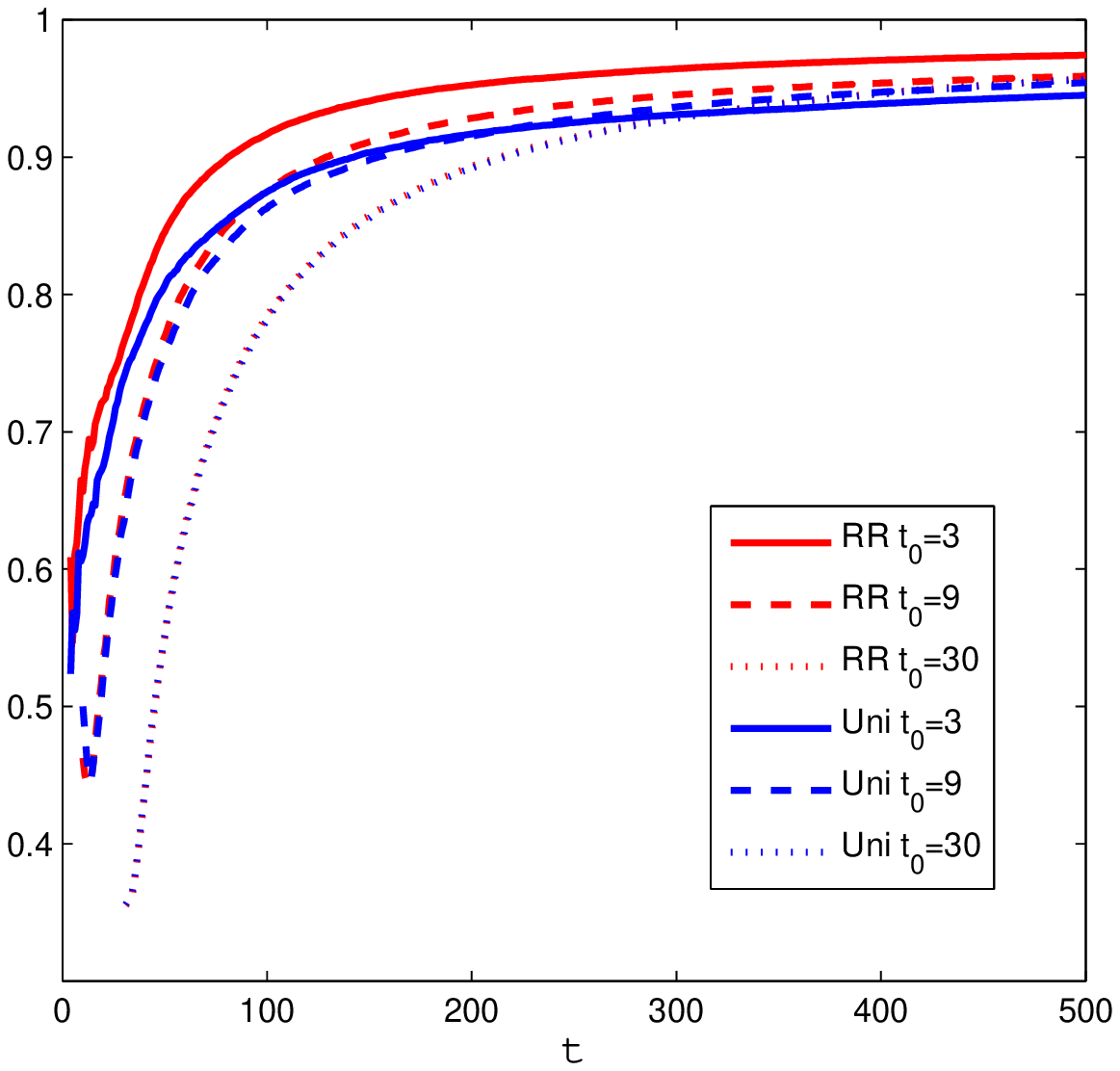}
\caption{The effect of the initial phase duration and initial strategy: tuned $\eps$-greedy algorithm.}
\label{fig:tunegreedyt0}
\end{figure}

\begin{figure}[t]
\centering
\includegraphics[width=8cm]{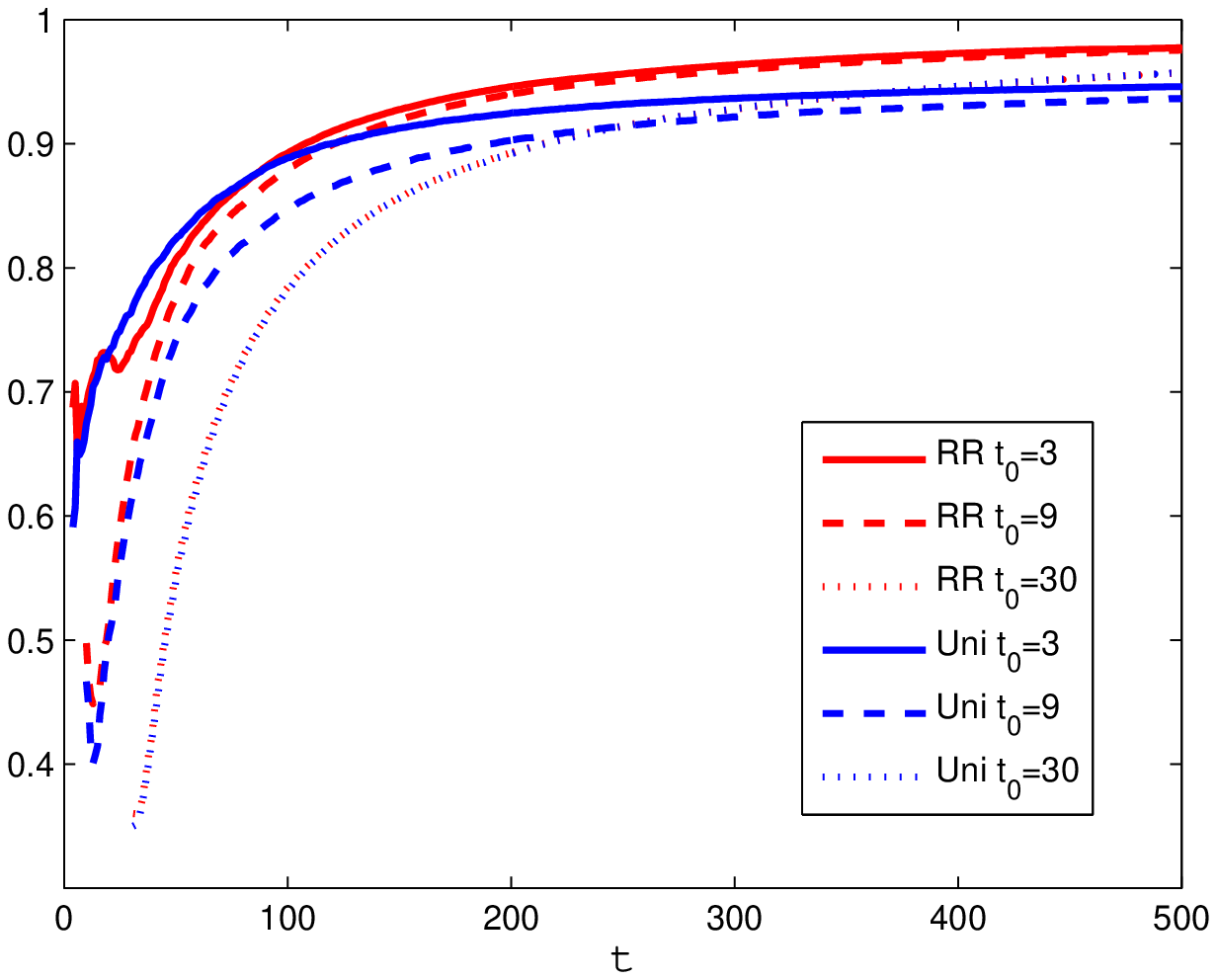}
\caption{The effect of the initial phase duration and initial strategy: UCB algorithm.}
\label{fig:ucbt0}
\end{figure}

In Figures~\ref{fig:greedyt0}-\ref{fig:ucbt0} for our numerical example
we plot the fraction of interests sent to the optimal arm for different durations ($t_0=3,9,30$)
of the initial phase for different algorithms with different initial phase strategies.

A bit surprisingly, it turns out that it is better to set up very short
duration of the initial phase.
Another important observation is that it is better to use the Round Robin initial strategy
rather than the uniformly random strategy. This is intuitively expected as by using the
Round Robin strategy we reduce the randomness.
Below we provide theoretical explanation of these phenomena.

The initial phase $[0,t_0-1]$ is characterized by large exploration effort. Here we would like to provide an estimate for the
period after which we can with high certainty rely on the choice of the best performing arm based on evaluated averages.
Specifically, let us estimate the probability of choosing the best arm (denoted by $*$) given the arms are chosen independently
before the end of the initialization phase.

Denote by $I_t$ the arm chosen at time slot $t$. Assume first that arms are chosen randomly and independently
during the initial phase with probability $p_j:=\Expectation[1\{I_t=j\}]$, $j=1,...,K$. In the case of uniformly random
strategy we have $p_j=1/K$. Let further $D$ be the maximum possible delay between choosing the arm and observing
the realization ($D = 1$ corresponds to no delay, i.e., receiving the chunk always in the slot immediately after
the slot when an interest was sent) and
$$
c_j:=D^2 + \frac{\Delta_j}{2}D + \frac{\Delta_j}{2} p_* D,
$$
where $\Delta_j=\mu_j-\mu_*$. Then, we have the following result.

\begin{theorem}\label{thm:Pbound}
If during the exploration phase we choose the arms randomly and independently with uniform distribution ($p_j=1/K$),
and at the end of the exploration period, at slot $t_0$, we choose the arm according to the estimated
average, the probability of choosing the best arm is lower bounded by
$$
\Probability[ \overline X_{*,T_*(t_0)} < \min_{j \neq *} \overline X_{j,T_j(t_0)} ]
$$
\begin{equation}\label{eq:bestarmprob}
\ge \prod_{j \neq *}
\left(1-\exp\left( -\frac{\Delta_j^2 (t_0-D)^2}{8K^2c_j^2t_0} \right)\right)^2
\end{equation}
\end{theorem}

A strong point of the above result is that the derived lower bound is given in terms
of exponential function, which means that starting from some value of $t_0$ the probability
of success will be very high. However, the bound (\ref{eq:bestarmprob}) can be loose.
Therefore, next we suggest an approximation of the success probability based on the
central limit theorem.

Also, it turns out that if the maximal delay is not too large, we do not introduce a large
error by considering only interests sent by the time $t_0-D$. Then, by the time $t_0$
we observe reply from all sent interests.

\begin{theorem}\label{thm:Pest}
If during the exploration phase we choose the arms randomly and independently with uniform distribution ($p_j=p_*=1/K$),
and if at the end of the exploration period, at slot $t_0$, we choose the arm according to the estimated
average, the probability of choosing the best arm can be approximated as follows:
$$
\Probability[ \overline X_{*,T_*(t_0-D)} < \min_{j \neq *} \overline X_{j,T_j(t_0-D)} ]
$$
$$
\approx \prod_{j \neq *}
\Phi \left( \frac{\Delta_j p_j \sqrt{t_0-D}}{2 \sqrt{p_j Var(X_j)+\Delta_j^2p_j(1-p_j)/4}} \right)
$$
\begin{equation}\label{eq:bestarmprobapprox}
\Phi \left( \frac{\Delta_j p_* \sqrt{t_0-D}}{2 \sqrt{p_* Var(X_*)+\Delta_j^2p_*(1-p_*)/4}} \right),
\end{equation}
where $\Phi(\cdot)$ is the cumulative distribution function of the standard normal random variable.
\end{theorem}

In the case when the Round Robin strategy is used in the initial phase, we can provide even sharper approximation.

\begin{theorem}\label{thm:PestRR}
If during the exploration phase we choose the arms according to the Round Robin strategy with
the first arm and the order chosen randomly with the uniform distribution,
and if at the end of the exploration period, at slot $t_0$, we choose the arm according to the estimated
average, the probability of choosing the best arm can be approximated as follows:
$$
\Probability[ \overline X_{*,T_*(t_0-D)} < \min_{j \neq *} \overline X_{j,T_j(t_0-D)} ]
$$
\begin{equation}\label{eq:bestarmprobapproxRR}
\approx \prod_{j \neq *}
\Phi \left( \Delta_j \sqrt{\frac{t_0-D}{3(Var(X_*)+Var(X_j))}} \right).
\end{equation}
\end{theorem}

\begin{figure}[t]
\centering
\includegraphics[width=8cm]{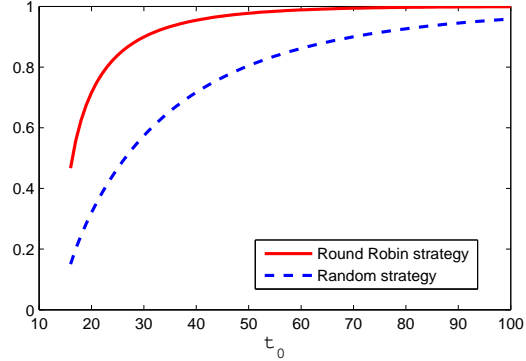}
\caption{Approximations for the probability of choosing the optimal arm at the end of the initial phase.}
\label{fig:CLTapprox}
\end{figure}

We consider now our numerical example with truncated negative binomial distributions with $D=15$.
In Figure~\ref{fig:CLTapprox} we plot the approximations (\ref{eq:bestarmprobapprox}) and (\ref{eq:bestarmprobapproxRR}),
which firstly confirm that it is enough to have a very short initial phase and secondly confirm our intuition
that the Round Robin strategy is better than the random strategy.

One may be interested in rough estimation of the number of time slots after which using estimated averages
the optimal arm will be selected with high probability. We can provide recommendation for such
value based on (\ref{eq:bestarmprobapproxRR}) and 2-sigma rule. If the arguments of the standard normal distribution
function are equal to two, then respective probabilities are greater than 0.977. Thus, we conclude that
after the time
\begin{equation}\label{eq:transapprox}
T \ge D + 12\frac{Var(X_*)+\max_j Var(X_j)}{\min_j \Delta_j^2},
\end{equation}
using the estimated averages and the RR strategy, we select the optimal arm
with probability at least $0.977^{K-1}$. In our numerical example, after 68 time slots the probability
of choosing correctly the optimal arm is estimated to be more than 0.95. This is even a conservative
estimation and in reality we need even shorter exploratory period.

\section{Logarithmic bound for the tuned $\eps$-greedy algorithm}
\label{sec:logbound}

In this section we finally prove that the regret (cumulative suboptimality) of employing the tuned $\eps$-greedy
algorithm is bounded logarithmically in $ t $, which is the same result as for the case without delay (and known to
be the best possible) \cite{ACF02}.



\begin{theorem}\label{thm:log}
Let $ a > 0 $ and $ 0 < d \le \min_{k : \mu_k > \mu_*} \Delta_k $, and let initial phase be run with the uniformly
random strategy. For all $ K > 1 $ and for all delay distributions $ F_{ 1 }, \dots, F_{ K } $ with support in $ [
1, D ] $, if algorithm tuned $ \eps $-greedy is run with input parameters $ t_0 > \eps_0 := a K / d^2 $, then the
probability that the algorithm chooses in slot $ t \ge t_{0} $ a suboptimal arm $ j $ is at most
\begin{align*}
&2 D \frac{ a }{ d^{ 2 } } \left( \ln \frac{ t d^{ 2 } e^{ 1 / 2 } }{ a K } \right) \left( \frac{ a K }{ t d^{ 2 } e^{ 1 / 2 } } \right)^{ \frac{ 3 a }{ 14 d^{ 2 } } } \\
&+ \frac{ 16 D^{ 3 } }{ d^{ 2 } } \exp \left\{ \frac{ D + 1 }{ 8 } \right\} \left( \frac{ a K }{ t d^{ 2 } e^{ 1 / 2 } } \right)^{ \frac{ a }{ 8 D^{ 2 } } } + \frac{ a }{ d^{ 2 } t }.
\end{align*}
\end{theorem}

This bound says that the cumulative probability of suboptimal decisions is logarithmic for $ a $ large enough
(surely if $ a > \max \{ 14 d^{ 2 } / 3, 8 D^2 \} $), because the instantaneous suboptimality at any slot $ t \ge
t_{ 0 } $ is of the order $ ( K - 1 ) a / d^{ 2 } t + o ( 1 / t ) $ for $ t \to \infty $.
We conclude that the smaller the number of arms (CCN neighbour routers) and the larger $d$, the difference between
the mean delays of the best and the strictly second-best arm, the better is the performance of the tuned $\eps$-greedy
algorithm.

\section{Conclusion}
\label{sec:conclusion}

The contribution of this paper is twofold. First, we have proposed tractable and well-performing interest forwarding algorithms
for CCN networks. We have demonstrated that the algorithms work fast and logarithmically
few interests are send suboptimally, which means that the resources of the user and CCN routers are efficiently managed.
Theoretical bounds show that the learning process is best achievable.

Second, we have also contributed to the theory of the multi-armed bandit problem with delayed information. This is an important
and challenging topic with few existing results. We have provided finite-time analysis of algorithms extended to this setting
and showed that the deterioration of their performance due to delays is not significant. Perhaps surprisingly, there is no need
to include a long exploratory phase, just a single datum from each arm is sufficient for an efficient performance of the
algorithms.

\section*{Acknowledgement}

We would like to thank Bruno Kauffmann, Luca Muscariello and Alain Simonian for stimulating discussions.


\appendix
\section{Appendix: Proofs}

\subsection{Auxiliary Material}

Let us state concentration inequalities to be used in the proofs of the theorems.
We first state the Chernoff-Hoeffding bound in a general form. This is called the Hoeffding's inequality in
\cite[p. 191]{Pollard1984}, citing \cite{Hoeffding1963}.

\begin{theorem}[Chernoff-Hoeffding bound]
Let $ Y_{ 1 }, Y_{ 2 }, \dots, Y_{ T } $ be independent random variables with zero means and bounded ranges $ a_{ t } \le Y_{ t
} \le b_{ t } $. Then, for each $ \eta > 0 $,
\begin{align*}
\Probability [ Y_{ 1 } + Y_{ 2 } + \dots + Y_{ T } \le - \eta ] &\le \exp \left\{ - 2 \eta^{ 2 } / \sum_{ t = 1 }^{ T } ( b_{ t } - a_{ t } )^{ 2 } \right\}
\\
\Probability [ Y_{ 1 } + Y_{ 2 } + \dots + Y_{ T } \ge \eta ] &\le \exp \left\{ - 2 \eta^{ 2 } / \sum_{ t = 1 }^{ T } ( b_{ t } - a_{ t } )^{ 2 } \right\}
\end{align*}
\end{theorem}


Let us state also the Bennett's inequality \cite{Bennett1962} and its consequence, the Bernstein's inequality.

\begin{theorem}[Bennett's inequality]
Let $ Y_{ 1 }, Y_{ 2 }, \dots, Y_{ T } $ be independent random variables with zero means and bounded ranges $ -M \le Y_{ t }
\le M $. Write $ \sigma_{ t }^{ 2 } $ for the variance of $ Y_{ t } $. Suppose $ V \ge \sigma_{ 1 }^{ 2 } + \dots +
\sigma_{ T }^{ 2 } $. Then, for each $ \eta > 0 $,
\begin{align*}
\Probability [ Y_{ 1 } + Y_{ 2 } + \dots + Y_{ T } \le - \eta ] &\le \exp \left\{ - \frac{ 1 }{ 2 } \eta^{ 2 } V^{ - 1 } B \left( M \eta V^{ - 1 } \right) \right\},
\\
\Probability [ Y_{ 1 } + Y_{ 2 } + \dots + Y_{ T } \ge \eta ] &\le \exp \left\{ - \frac{ 1 }{ 2 } \eta^{ 2 } V^{ - 1 } B \left( M \eta V^{ - 1 } \right) \right\},
\end{align*}
where $ B ( \lambda ) := 2 \lambda^{ - 2 } [ ( 1 + \lambda ) \log ( 1 + \lambda ) - \lambda ] $, for $ \lambda > 0 $.
\end{theorem}


According to \cite[p. 193]{Pollard1984}:

\begin{quote}
``The function $ B ( \cdot ) $ is well-behaved: continuous, decreasing, and $ B ( 0+ ) = 1 $. When $ \lambda $ is large, $ B(
\lambda ) \approx 2 \lambda^{ - 1 } \log \lambda $ in the sense that the ratio tends to one as $ \lambda \to \infty $; the
Bennett Inequality does not give a true exponential bound for $ \eta $ compared to $ V / M $. For smaller $ \eta $ it comes
very close to the bound for normal tail probabilities. Problem 2 shows that $ B( \lambda ) \ge ( 1 + \frac{ 1 }{ 3 } \lambda
)^{ - 1 } $ for all $ \lambda > 0 $.''
\end{quote}

Using the last bound, we get the Bernstein's inequality.

\begin{theorem}[Bernstein's inequality]
Let $ Y_{ 1 }, Y_{ 2 }, \dots, Y_{ T } $ be independent random variables with zero means and bounded ranges $ -M \le Y_{ t }
\le M $. Write $ \sigma_{ t }^{ 2 } $ for the variance of $ Y_{ t } $. Suppose $ V \ge \sigma_{ 1 }^{ 2 } + \dots +
\sigma_{ T }^{ 2 } $. Then, for each $ \eta > 0 $,
\begin{align*}
\Probability [ Y_{ 1 } + Y_{ 2 } + \dots + Y_{ T } \le - \eta ] &\le \exp \left\{ - \frac{ 1 }{ 2 } \eta^{ 2 } / \left( V + \frac{ 1 }{ 3 } M \eta \right) \right\},
\\
\Probability [ Y_{ 1 } + Y_{ 2 } + \dots + Y_{ T } \ge \eta ] &\le \exp \left\{ - \frac{ 1 }{ 2 } \eta^{ 2 } / \left( V + \frac{ 1 }{ 3 } M \eta \right) \right\}.
\end{align*}
\end{theorem}

Finally, we present the Azuma's inequality.

\begin{theorem}[Azuma's inequality]
Let $Z_t$ be a martingale with zero mean and bounded increment, i.e.,
$$
|Z_t-Z_{t-1}| \le c(t),
$$
almost surely. Then, for all positive integers $t$ and all positive reals $\lambda$, we have
$$
P[Z_t \ge \lambda] \le \exp \left( -\frac{\lambda^2}{2\sum_{s=1}^t c^2(s)} \right).
$$
\end{theorem}

\subsection{Proof of Theorem \ref{thm:Pbound}}
We need to evaluate the following probability:

$$
P[ \bar X_{*,T_*(t_0)} < \min_{j \neq *} \bar X_{j,T_j(t_0)}]
=P[\cap_{j \neq *} \{ \bar X_{*,T_*(t_0)} < \bar X_{j,T_j(t_0)} \} ]
$$
$$
=\prod_{j \neq *} P[ \bar X_{*,T_*(t_0)} < \bar X_{j,T_j(t_0)} ]
$$
$$
\ge \prod_{j \neq *} P[ \{ \bar X_{*,T_*(t_0)} < \mu_* + \frac{\Delta_j}{2} \}
\cap \{\bar X_{j,T_j(t_0)} \ge \mu_j - \frac{\Delta_j}{2} \} ]
$$
\begin{equation}\label{eq:errorprob}
= \prod_{j \neq *} P[ \bar X_{*,T_*(t_0)} < \mu_* + \frac{\Delta_j}{2} ]
P[ \bar X_{j,T_j(t_0)} \ge \mu_j - \frac{\Delta_j}{2} ].
\end{equation}

Now let us estimate the probability $P[ \bar X_{*,T_*(t_0)} < \mu_* + \frac{\Delta_j}{2} ]$.

$$
P[ \bar X_{*,T_*(t_0)} < \mu_* + \frac{\Delta_j}{2} ]
= 1 - P[ \bar X_{*,T_*(t_0)} \ge \mu_* + \frac{\Delta_j}{2} ]
$$
$$
=1-P\left[ \frac{\sum_{s=1}^{t_0} 1\{I_s=*\} X_*(s) 1\{s+X_*(s) \le t_0 \} }{\sum_{s=1}^{t_0} 1\{I_s=*\} 1\{s+X_*(s) \le t_0 \}}
\ge \mu_* + \frac{\Delta_j}{2}\right]
$$
$$
=1-P\left[ \sum_{s=1}^{t_0} 1\{I_s=*\} (X_*(s)-\mu_*) 1\{s+X_*(s) \le t_0 \}
\ge \frac{\Delta_j}{2}\sum_{s=1}^{t_0} 1\{I_s=*\} 1\{s+X_*(s) \le t_0 \} \right]
$$

$$
=1-P\left[ \sum_{s=1}^{t_0} 1\{I_s=*\} (X_*(s)-\mu_*) 1\{s+X_*(s) \le t_0 \}\right.
$$
$$\left.
-\frac{\Delta_j}{2}\sum_{s=1}^{t_0} (1\{I_s=*\}-p_*) 1\{s+X_*(s) \le t_0 \}
\ge \frac{\Delta_j}{2}p_* \sum_{s=1}^{t_0} 1\{s+X_*(s) \le t_0 \} \right]
$$

$$
=1-P\left[ \sum_{s=1}^{t_0} 1\{I_s=*\} (X_*(s)-\mu_*) 1\{s+X_*(s) \le t_0 \}\right.
$$
$$
-\frac{\Delta_j}{2}\sum_{s=1}^{t_0} (1\{I_s=*\}-p_*) 1\{s+X_*(s) \le t_0 \}
-\frac{\Delta_j}{2}p_* \sum_{s=1}^{t_0} (1\{s+X_*(s) \le t_0 \}-q_{*,t_0-s})
$$
$$\left.
\ge \frac{\Delta_j}{2}p_* (t_0-D+\sum_{i=1}^{D}q_{*,i}) \right],
$$
where $q_{*,i} := P[X_*(t) \le i]$.

Next we define
$$
Z_{j,t} := \sum_{s=1}^t 1\{I_s=*\} (X_*(s)-\mu_*) 1\{s+X_*(s) \le t \}
$$
$$
-\frac{\Delta_j}{2}\sum_{s=1}^t (1\{I_s=*\}-p_*) 1\{s+X_*(s) \le t \}
$$
$$
-\frac{\Delta_j}{2}p_* \sum_{s=1}^t (1\{s+X_*(s) \le t \}-q_{*,t-s}).
$$

It is a martingale (with respect to the sequence of the observed delays) with zero mean and
bounded increment
$$
|Z_t - Z_{t-1}| \le c_j,
$$
with $c_j=D^2 + \frac{\Delta_j}{2}D + \frac{\Delta_j}{2} p_* D$.

Thus, we can apply Azuma's inequality for martingales, which gives in our case
$$
P[ \bar X_{*,T_*(t_0)} < \mu_* + \frac{\Delta_j}{2} ]
\ge 1-\exp\left(-\frac{\Delta_j^2/4 p_*^2 (t_0-D+\sum_{i=1}^D q_{*,i})^2}{2c^2_jt_0} \right)
$$
\begin{equation}\label{eq:Azumaj}
\ge 1-\exp\left(-\frac{\Delta_j^2/4 p_*^2 (t_0-D)^2}{2c^2_jt_0} \right).
\end{equation}
Similarly, we have
\begin{equation}\label{eq:Azuma*}
P[ \bar X_{j,T_j(t_0)} \ge \mu_j - \frac{\Delta_j}{2} ]
\ge 1-\exp \left( -\frac{\Delta_j^2/4 p_j^2 (t_0-D)^2}{2c^2_jt} \right).
\end{equation}

Substituting (\ref{eq:Azumaj}) and (\ref{eq:Azuma*}) into (\ref{eq:errorprob}), we
complete the proof.

\subsection{Proof of Theorem \ref{thm:Pest}}

Similarly to (\ref{eq:errorprob}), we have
$$
P[ \bar X_{*,T_*(t_0-D)} < \min_{j \neq *} \bar X_{j,T_j(t_0-D)}]
$$
\begin{equation}\label{eq:errorprob1}
\ge \prod_{j \neq *} P[ \bar X_{*,T_*(t_0-D)} < \mu_* + \frac{\Delta_j}{2} ]
P[ \bar X_{j,T_j(t_0-D)} \ge \mu_j - \frac{\Delta_j}{2} ]
\end{equation}

Define
$$
Y_t=\sum_{s=1}^t \left(1\{I_s=*\}(X_{*,s}-\mu_*) - \frac{\Delta_j}{2}(1\{I_s=*\}-p_*)\right).
$$
Then, we can use the Central Limit theorem to estimate the probability
$$
P[ \bar X_{*,T_*(t_0-D)} < \mu_* + \frac{\Delta_j}{2} ]
= P[ Y_{t_0-D} < \frac{\Delta_j}{2} p_* (t_0-D)]
$$
$$
= P[ \frac{Y_{t_0-D}}{\sqrt{(t_0-D)(p_* Var(X_*)+\Delta_j^2p_*(1-p_*)/4)}} <
$$
$$
\frac{\Delta_j p_* (t_0-D)}{2 \sqrt{(t_0-D)(p_* Var(X_*)+\Delta_j^2p_*(1-p_*)/4)}}],
$$
which gives
\begin{equation}\label{eq:cltj}
P[ \bar X_{*,T_*(t_0-D)} < \mu_* + \frac{\Delta_j}{2} ]
\approx \Phi \left( \frac{\Delta_j p_* \sqrt{t_0-D}}{2 \sqrt{p_* Var(X_*)+\Delta_j^2p_*(1-p_*)/4}} \right),
\end{equation}
where $\Phi(\cdot)$ is the standard normal distribution function.
Similarly, we obtain
\begin{equation}\label{eq:clt*}
P[ \bar X_{j,T_j(t_0-D)} \ge \mu_j - \frac{\Delta_j}{2} ]
\approx \Phi \left( \frac{\Delta_j p_j \sqrt{t_0-D}}{2 \sqrt{p_j Var(X_j)+\Delta_j^2p_j(1-p_j)/4}} \right).
\end{equation}
The substitution of (\ref{eq:cltj}) and (\ref{eq:clt*}) into (\ref{eq:errorprob1}) yields the result.

\bigskip

The proof of Theorem~3 is simpler than the proof of Theorem~2 and it is omitted.

\subsection{Proof of Theorem \ref{thm:log}}

Note that the assumption $ t \ge t_0 $ means that we are in the exploitation phase, and let us denote by $ \eps_t
:= \eps_0 / t $ for all $ t \ge t_0 $, while $ \eps_t := 1 $ for all $ t < t_0 $.

Let $ \overline{ X }_{ j, s } $ be the sample mean of observed delays (costs) if arm $ j $ was chosen $ s $ times
conditioned on the delay distribution. Let $ \overline{ X }_{ j, s, u } $ be the sample mean of observed delays if
arm $ j $ was chosen $ s $ times having obtained $ u \le s $ observations. Let $ S_{ j } ( t ) $ denote the number
of times arm $ j $ was chosen in the first $ t $ slots $ [ 0, t - 1 ] $. Recall that $ I_{ t } $ denotes the arm
chosen at slot $ t $. Then we have
\begin{align*}
\Probability \left[ I_{ t } = j \right] &\le ( 1 - \eps_{ t } ) \Probability \left[ \overline{ X }_{ j, S_{ j } ( t ) } \le \max_{ k \neq j } \overline{ X }_{ k, S_{ k } ( t ) } \right] + \frac{ \eps_{ t } }{ K }.
\end{align*}
Note that here we have an inequality in order to account for an arbitrary rule of breaking ties in deciding the arm
to choose in case several arms have the same lowest sample mean.

If $ j \neq * $ (where $ * $ denotes any of the best arms), then we can bound it by
\begin{align}
\Probability \left[ I_{ t } = j \right] &\le \Probability \left[ \overline{ X }_{ j, S_{ j } ( t ) } \le \overline{ X }_{ *, S_{ * } ( t ) } \right] + \frac{ \eps_{ t } }{ K } \notag \\
&\le \Probability \left[ \overline{ X }_{ j, S_{ j } ( t ) } \le \mu_{ j } - \frac{ \Delta_{ j } }{ 2 } \right] + \Probability \left[ \overline{ X }_{ *, S_{ * } ( t ) } \ge \mu_{ * } + \frac{ \Delta_{ j } }{ 2 } \right] + \frac{ \eps_{ t } }{ K }. \label{eq:A}
\end{align}

Let now $ U_{ j, s } ( t ) $ denote the number of observed realizations by the beginning of slot $ t $ from arm $ j
$ given that it was chosen $ s $ times in the slots $ [ 0, t - 1 ] $. In order to upperbound the first two terms in
\eqref{eq:A} (by an expression independent of $ j $), let us study the following expression next.
\begin{align}
\Probability &\left[ \overline{ X }_{ j, S_{ j } ( t ) } \ge \mu_{ j } + \frac{ \Delta_{ j } }{ 2 } \right] = \sum_{ s = 1 }^{ t } \Probability \left[ S_{ j } ( t ) = s \text{ and } \overline{ X }_{ j, s } \ge \mu_{ j } + \frac{ \Delta_{ j } }{ 2 } \right] \notag \\
&= \sum_{ s = 1 }^{ t } \Probability \left[ S_{ j } ( t ) = s \ | \ \overline{ X }_{ j, s } \ge \mu_{ j } + \frac{ \Delta_{ j } }{ 2 } \right] \Probability \left[ \overline{ X }_{ j, s } \ge \mu_{ j } + \frac{ \Delta_{ j } }{ 2 } \right] \notag \\
&= \sum_{ s = 1 }^{ t } \Probability \left[ S_{ j } ( t ) = s \ | \ \overline{ X }_{ j, s } \ge \mu_{ j } + \frac{ \Delta_{ j } }{ 2 } \right] \sum_{ u = 1 }^{ s } \Probability \left[ U_{ j, s } ( t ) = u \text{ and } \overline{ X }_{ j, s, u } \ge \mu_{ j } + \frac{ \Delta_{ j } }{ 2 } \right] \notag \\
&= \sum_{ s = 1 }^{ t } \Probability \left[ S_{ j } ( t ) = s \ | \ \overline{ X }_{ j, s } \ge \mu_{ j } + \frac{ \Delta_{ j } }{ 2 } \right] \sum_{ u = 1 }^{ s } \Probability \left[ U_{ j, s } ( t ) = u \ | \ \overline{ X }_{ j, s, u } \ge \mu_{ j } + \frac{ \Delta_{ j } }{ 2 } \right] \Probability \left[ \overline{ X }_{ j, s, u } \ge \mu_{ j } + \frac{ \Delta_{ j } }{ 2 } \right]. \label{eq:C}
\end{align}

Assuming that $ \Probability \left[ \overline{ X }_{ j, s, u } \ge \mu_{ j } + \frac{ \Delta_{ j } }{ 2 } \right] > 0
$, then, for $ 1 \le u \le s $,
\begin{align*}
\Probability \left[ U_{ j, s } ( t ) = u \ | \ \overline{ X }_{ j, s, u } \ge \mu_{ j } + \frac{ \Delta_{ j } }{ 2 } \right]
\begin{cases}
  = 0, & \text{ if } s - D + 1 > u, \\
  \le 1, & \text{ if } s - D + 1 \le u, \\
\end{cases}
\end{align*}
because there can be at most $ D - 1 $ unobserved realizations of the chosen arms ($ s - u \le D - 1 $). Hence,
\begin{align*}
&\sum_{ u = 1 }^{ s } \Probability \left[ U_{ j, s } ( t ) = u \ | \ \overline{ X }_{ j, s, u } \ge \mu_{ j } + \frac{ \Delta_{ j } }{ 2 } \right] \Probability \left[ \overline{ X }_{ j, s, u } \ge \mu_{ j } + \frac{ \Delta_{ j } }{ 2 } \right] \\
&\le \sum_{ u = \max\{ 1, s - D + 1 \} }^{ s } \Probability \left[ ( \overline{ X }_{ j, s, u } - \mu_{ j } ) u \ge \frac{ \Delta_{ j } u }{ 2 } \right] \\
&\le \sum_{ u = \max\{ 1, s - D + 1 \} }^{ s } \exp \left\{ - 2 \left( \frac{ \Delta_{ j } u }{ 2 } \right)^{ 2 } / u \left( 2 D \right)^{ 2 } \right\} = \sum_{ u = \max\{ 1, s - D + 1 \} }^{ s } \exp \left\{ - \left( \frac{ \Delta_{ j }^{ 2 } u }{ 8 D^{ 2 } } \right) \right\},
\end{align*}
where the last inequality is due to the Chernoff-Hoeffding bound (employed with $ \eta = \frac{ \Delta_{ j } u }{ 2 }, b_{ t } = D, a_{ t } = -D, T = u $).

Upperbounding the last geometric sum by a sum of constants equal to the first term, we further have
\begin{align*}
&\sum_{ u = 1 }^{ s } \Probability \left[ U_{ j, s } ( t ) = u \ | \ \overline{ X }_{ j, s, u } \ge \mu_{ j } + \frac{ \Delta_{ j } }{ 2 } \right] \Probability \left[ \overline{ X }_{ j, s, u } \ge \mu_{ j } + \frac{ \Delta_{ j } }{ 2 } \right] \\
&\le D \exp \left\{ - \frac{ \Delta_{ j }^{ 2 } }{ 8 D^{ 2 } } \max \{ 1, s - D + 1 \} \right\}.
\end{align*}

This bound plugged into \eqref{eq:C} therefore gives us
\begin{align}
\Probability &\left[ \overline{ X }_{ j, S_{ j } ( t ) } \ge \mu_{ j } + \frac{ \Delta_{ j } }{ 2 } \right] \notag \\
&\le D \sum_{ s = 1 }^{ t } \Probability \left[ S_{ j } ( t ) = s \ | \ \overline{ X }_{ j, s } \ge \mu_{ j } + \frac{ \Delta_{ j } }{ 2 } \right] \exp \left\{ - \frac{ \Delta_{ j }^{ 2 } }{ 8 D^{ 2 } } \max \{ 1, s - D + 1 \} \right\} \notag \\
&\le D \sum_{ s = 1 }^{ \infty } \Probability \left[ S_{ j } ( t ) = s \ | \ \overline{ X }_{ j, s } \ge \mu_{ j } + \frac{ \Delta_{ j } }{ 2 } \right] \exp \left\{ - \frac{ \Delta_{ j }^{ 2 } }{ 8 D^{ 2 } } \max \{ 1, s - D + 1 \} \right\} \notag \\
&\le D \exp \left\{ - \frac{ \Delta_{ j }^{ 2 } }{ 8 D^{ 2 } } \right\} \sum_{ s = 1 }^{ D - 1 } \Probability \left[ S_{ j } ( t ) = s \ | \ \overline{ X }_{ j, s } \ge \mu_{ j } + \frac{ \Delta_{ j } }{ 2 } \right] \notag \\
&+ D \sum_{ s = D }^{ \lfloor E \rfloor } \Probability \left[ S_{ j } ( t ) = s \ | \ \overline{ X }_{ j, s } \ge \mu_{ j } + \frac{ \Delta_{ j } }{ 2 } \right] \exp \left\{ - \frac{ \Delta_{ j }^{ 2 } }{ 8 D^{ 2 } } ( s - D + 1 ) \right\} \notag \\
&+ D \sum_{ s = \lfloor E \rfloor + 1 }^{ \infty } \Probability \left[ S_{ j } ( t ) = s \ | \ \overline{ X }_{ j, s } \ge \mu_{ j } + \frac{ \Delta_{ j } }{ 2 } \right] \exp \left\{ - \frac{ \Delta_{ j }^{ 2 } }{ 8 D^{ 2 } } ( s - D + 1 ) \right\} \label{eq:B}
\end{align}
where
\begin{align*}
E := \frac{ 1 }{ 2 K } \sum_{ s = 0 }^{ t - 1 } \eps_{ s }.
\end{align*}


Note that if $ \lfloor E \rfloor \ge D - 1 $, then the above decomposition of the sum in the last step in fact
holds as equality. In case $ \lfloor E \rfloor < D - 1 $, the second term is zero and some of the summands appear
both in the first and in the third term, therefore the inequality holds.

The sum of the first and second terms in \eqref{eq:B} can be upperbounded by
\begin{align*}
D \sum_{ s = 1 }^{ \lfloor E \rfloor } \Probability \left[ S_{ j } ( t ) = s \ | \ \overline{ X }_{ j, s } \ge \mu_{ j } + \frac{ \Delta_{ j } }{ 2 } \right]
\end{align*}
omitting the exponential terms ($ \le 1 $), which is further upperbounded (as in \cite{ACF02}) by
\begin{align*}
D \sum_{ s = 1 }^{ \lfloor E \rfloor } \Probability \left[ S_{ j }^{ \text{R} } ( t ) \le s \ | \ \overline{ X }_{ j, s } \ge \mu_{ j } + \frac{ \Delta_{ j } }{ 2 } \right] \le D E \Probability \left[ S_{ j }^{ \text{R} } ( t ) \le E \right],
\end{align*}
where $ S_{ j }^{ \text{R} } ( t ) \le S_{ j } ( t ) $ is the the number of times arm $ j $ was chosen in the first
$ t $ slots $ [ 0, t - 1 ] $ at random. Using the Bernstein inequality (with $ Y_{ s + 1 } $ for $ s = 0, 1, \dots,
t - 1 $ being the random variable of sending the interest to router $ j $ at slot $ s $, with expected value $
\eps_{ s } / K $, bounded by $ M = 1 $, and variance $ \sigma_{ s + 1 }^{ 2 } = ( 1 - \eps_{ s } / K ) ( 0 - \eps_{
s } / K )^{ 2 } + \eps_{ s } / K ( 1 - \eps_{ s } / K )^{ 2 } = ( 1 - \eps_{ s } / K ) \eps_{ s } / K \le \eps_{ s
} / K $, so that $ V = 2 E $, and taking $ \eta = E $), we have (a slightly tighter upperbound than in
\cite{ACF02})
\begin{align*}
\Probability \left[ S_{ j }^{ \text{R} } ( t ) \le E \right] \le \exp \left\{ - \frac{ 3 }{ 14 } E \right\}
\end{align*}
and for $ t \ge a K / d^{ 2 } $, we lowerbound $ E $ as in \cite{ACF02} (denoted $ x_{ 0 } $ there),
\begin{align}
E \ge \frac{ a }{ d^{ 2 } } \ln \frac{ t d^{ 2 } e^{ 1 / 2 } }{ a K }. \label{eq:D}
\end{align}
Therefore, the sum of the first and second terms in \eqref{eq:B} can be upperbounded by
\begin{align*}
D \frac{ a }{ d^{ 2 } } \left( \ln \frac{ t d^{ 2 } e^{ 1 / 2 } }{ a K } \right) \left( \frac{ a K }{ t d^{ 2 } e^{ 1 / 2 } } \right)^{ \frac{ 3 a }{ 14 d^{ 2 } } }.
\end{align*}

As in \cite{ACF02}, the third term in \eqref{eq:B} can be upperbounded by
\begin{align*}
\frac{ 8 D^{ 3 } }{ \Delta_{ j }^{ 2 } } \exp \left\{ - \frac{ \Delta_{ j }^{ 2 } }{ 8 D^{ 2 } } ( \lfloor E \rfloor - D ) \right\} = \frac{ 8 D^{ 3 } }{ \Delta_{ j }^{ 2 } } \exp \left\{ \frac{ \Delta_{ j }^{ 2 } }{ 8 D^{ 2 } } D \right\} \exp \left\{ - \frac{ \Delta_{ j }^{ 2 } }{ 8 D^{ 2 } } \lfloor E \rfloor \right\}
\end{align*}
omitting the probability term ($ \le 1 $) and using $ \displaystyle \sum_{ s = r + 1 }^{ \infty } e^{ - \alpha s }
\le \frac{ 1 }{ \alpha  } e^{ - \alpha r } $, with $ r = \lfloor E \rfloor - D , \alpha = \frac{ \Delta_{ j }^{ 2 }
}{ 8 D^{ 2 } } $. Further, using $ \lfloor E \rfloor \ge E - 1 $, this can be upperbounded by
\begin{align*}
\frac{ 8 D^{ 3 } }{ \Delta_{ j }^{ 2 } } \exp \left\{ \frac{ \Delta_{ j }^{ 2 } ( D + 1 ) }{ 8 D^{ 2 } } \right\} \exp \left\{ - \frac{ \Delta_{ j }^{ 2 } }{ 8 D^{ 2 } } E \right\}
\end{align*}
and further by
\begin{align*}
\frac{ 8 D^{ 3 } }{ d^{ 2 } } \exp \left\{ \frac{ D^{ 2 } ( D + 1 ) }{ 8 D^{ 2 } } \right\} \left( \frac{ a K }{ t d^{ 2 } e^{ 1 / 2 } } \right)^{ \frac{ a }{ 8 D^{ 2 } } }.
\end{align*}
where the bound for the third term is obtained using \eqref{eq:D}.

So, we have
\begin{align*}
\Probability &\left[ \overline{ X }_{ j, S_{ j } ( t ) } \ge \mu_{ j } + \frac{ \Delta_{ j } }{ 2 } \right] \le D \frac{ a }{ d^{ 2 } } \left( \ln \frac{ t d^{ 2 } e^{ 1 / 2 } }{ a K } \right) \left( \frac{ a K }{ t d^{ 2 } e^{ 1 / 2 } } \right)^{ \frac{ 3 a }{ 14 d^{ 2 } } } \\
&+ \frac{ 8 D^{ 3 } }{ d^{ 2 } } \exp \left\{ \frac{ D + 1 }{ 8 } \right\} \left( \frac{ a K }{ t d^{ 2 } e^{ 1 / 2 } } \right)^{ \frac{ a }{ 8 D^{ 2 } } }.
\end{align*}

In fact, the same upperbound holds for $ \Probability \left[ \overline{ X }_{ *, S_{ * } ( t ) } \ge \mu_{ * } + \frac{
\Delta_{ j } }{ 2 } \right] $, which is the second term in \eqref{eq:A}.

Finally, we have $ \eps_{ t } = a K / d^{ 2 } t $ to plug in the third term in \eqref{eq:A}, therefore
\begin{align*}
\Probability \left[ I_{ t } = j \right] &\le 2 D \frac{ a }{ d^{ 2 } } \left( \ln \frac{ t d^{ 2 } e^{ 1 / 2 } }{ a K } \right) \left( \frac{ a K }{ t d^{ 2 } e^{ 1 / 2 } } \right)^{ \frac{ 3 a }{ 14 d^{ 2 } } } + \frac{ 16 D^{ 3 } }{ d^{ 2 } } \exp \left\{ \frac{ D + 1 }{ 8 } \right\} \left( \frac{ a K }{ t d^{ 2 } e^{ 1 / 2 } } \right)^{ \frac{ a }{ 8 D^{ 2 } } } + \frac{ a }{ d^{ 2 } t }.
\end{align*}

\newpage
\tableofcontents

\end{document}